# Comments on the observation of quantum oscillations in cuprates

H.D. Drew

*Center for Nanophysics and Advanced Materials, Department of Physics, University of Maryland, College Park, Maryland 20742, USA.*

Recently the observation of quantum oscillations was reported in YBCO (ortho-II YBa$_2$Cu$_3$O$_{6.5}$ and YBa$_2$Cu$_4$O$_8$) with periods in 1/H indicative of Fermi surface pockets much smaller than the large hole pocket in optimally doped cuprates observed by ARPES.[1,2] This is an important development in cuprate physics. If their interpretation is correct, the existence of small Fermi pockets in these materials will have profound implications on the structure of the pseudogap state. It heralds a new era in which cuprate Fermiology can be used as a powerful tool to critically address the different ideas about the exotic properties of these materials and, of course, the mechanism for superconductivity. These measurements were made in high magnetic fields on oxygen ordered single crystals of YBCO to achieve the carrier long lifetimes $\tau$ and high cyclotron frequency $\omega_C$ needed to achieve sufficiently large $\omega_C\tau$ for oscillations to be observable. This requirement may limit the applicability of the technique to YBCO at a few doping levels using the materials and magnetic fields currently available. These requirements are not so severe with other techniques that also have given evidence, more indirectly, for small Fermi pockets in both electron and hole doped cuprates. These include ARPES,[3,4] DC magneto-transport and Hall effect measured at infrared (IR) frequencies.[5-8] Nevertheless, the observation of quantum oscillations is compelling and promises to put Fermiology on a firm footing and decisively on the agenda for the cuprates. In this comment I address some of the issues associated with this new development.

Evidence for small Fermi pocket has also been seen in ARPES measurements on underdoped cuprates in the form of "Fermi arcs".[3,4] The k-space locations of these arcs are associated with the magnetic Brillouin zone of the Néel state of the parent compounds. In a Fermi pocket scenario the Fermi arcs are one half of a Fermi pocket and the other half is not observable in photoemission in zone folding models. Correlating ARPES data on Fermi arcs with the quantum oscillation data would be highly desirable but is hindered by the scarcity of ARPES data on YBCO due to materials problems [ref. 3 and refs. therein].

IR Hall measurements on underdoped La$_{2-x}$Sr$_x$CuO$_4$ (LSCO) and YBa$_2$Cu$_3$O$_7$ (YBCO) have also given evidence for Fermi surface reconstruction and formation of small Fermi pockets.[6,7] The measured IR Hall angle, given by $\theta_H(\omega) = \omega_H/(\gamma - i\omega)$, allows a separate determination of the Hall frequency[9], $\omega_H \equiv eB/m_H c$, and the Hall scattering rate $\gamma$. In the experiments $\omega_H$ was observed to increase rapidly as doping was reduced from optimal. The observed increase in $\omega_H$ implies a reduction in the Hall mass, m$_H$. This is inconsistent with the expected behavior at the approach to a Mott state in which the effective mass diverges.[10-12] The IR Hall results, however, are consistent with existence of small Fermi pockets such as expected from the formation of density wave states such as spin density waves (SDW) or d-density waves (DDW) due to the corresponding reconstruction of the Fermi surface.[13,14]

More recently we reported additional evidence for SDW gap effects in the underdoped electron doped cuprate $Pr_{2-x}Ce_xCuO_4$ (PCCO) from optical and IR Hall measurements.[8,15] Optical transitions across a SDW gap produce signatures in both $\sigma_{xx}$[15] and, more dramatically and to higher doping (x) in $\sigma_{xy}$ at mid IR frequencies.[8] Also, the DC Hall effect and magnetoresistance show effects consistent with the SDW scenario.[5] At low temperatures $R_H$ switches from positive for x>0.16 to negative for x<0.16 and grows large as the doping is further reduced. This behavior is consistent with a large hole Fermi surface at high electron doping reconstructing to small electron pockets in a SDW state below x=0.16. The reconstructed Fermi surface produces electron pockets near (π,0) that shrink as x is reduced. Above $x = 0.16$ the optical and IR and DC Hall response reverts to a Drude-like behavior with no evidence for SDW effects.[8,15] Similar optical and transport signatures are expected for the DDW state.[13]

Therefore, the mid IR Hall measurements give two kinds of evidence for a density wave ground state in underdoped cuprates and, the observations in electron and hole doped cuprates are strikingly different. In h-cuprates we see evidence for small Fermi pockets which is the signature of a partial gapping of the Fermi surface.[6,7] In e-doped cuprates we see direct evidence for a DW gap.[8,15] In both cases, however, the IR Hall results support the formation of small Fermi pockets in the underdoped cuprates.

To confirm the identification of the quantum oscillations observed in YBCO several important issues remain to be resolved. One is the presence of $CuO_2$ chains in YBCO compounds which may confound the interpretation. The chains produce quasi 1D electronic bands in addition to the quasi 2D bands produced by the $CuO_2$ planes that characterize all the cuprate metals and where the superconductivity resides. One possibility is that the observed oscillations arise from small pockets associated with the chain bands due to their hybridization with the plane bands. This issue was discussed in the quantum oscillation reports[1,2] and in a subsequent publication[16] and remains a critical question to be resolved. There are other potential sources of oscillations coming from the chain bands. It is well known from work on organic metals that quasi 1D bands can support magnetic field induced spin density wave states that can appear very similar to the quantum oscillations produced by a closed Fermi surface.[17,18] In addition, oscillatory effects arising from other effects have been observed and explained in quasi 1D organic metals.[18,19] Addressing the alternative interpretations is an important priority.

If the chain bands are not responsible for the oscillations, then small Fermi pockets may be a general feature of the underdoped cuprates as is also suggested by the ARPES and IR hall results. What is their origin? The SDW scenario is a natural mechanism for producing pockets by the coherent scattering by zone corner (π,π) magnons producing a gapping of the Fermi surface at the magnetic Brillouin zone.[8,13,14] This picture is consistent with the ARPES observations. However, it is generally expected that long range magnetic order is a necessary condition for Fermi surface reconstruction which has not been observed in the doping range where Fermi arcs occur or at the hole doping values corresponding to the quantum oscillations experiments. In hole doped cuprates long range antiferromagnetic order disappears at low doping. In $La_{2-x}Sr_xCuO_4$, for example, AFM order disappears above x=0.02. The situation for electron doped cuprates, however, is not so clear. The magneto-optical and transport SDW signatures occur both below and above the critical doping level where long range antiferromagnetic order is observed in PCCO in neutron scattering.[20] The neutron data[20] indicates long range AFM order up to x = 0.13 while various experiment suggest SDW gapping for doping up to x = 0.16.[5,8,15] Similarly, the DDW scenario, while not requiring long range AFM order, does require a phase transition which has not yet been observed. Therefore, while the DW scenarios look promising as mechanisms for the formation of small Fermi pockets they raise questions that may also be the critical questions for an understanding the pseudogap state.

A surprising result in the quantum oscillations papers is the large value of the cyclotron effective mass considering the small Fermi pockets inferred from the experiments (3% of the Brillouin zone area). They find $m_C \approx 3 m_0$ which is close to the band value for the unreconstructed Fermi surface.[1,2] In general, a small Fermi pocket implies a rapid variation of the energy dispersion resulting in a small mass. In this regard the IR Hall results and the quantum oscillations results appear to be inconsistent. The IR Hall data indicates a rapid decrease of the Hall mass with underdoping.[6,7] These two masses, $m_C$ and $m_H$ are related but not equal in general.[9] For example, for an elliptical Fermi surface $m_C = \sqrt{m_1 m_2}$, where $m_1$ and $m_2$ are the two principle components of the mass tensor. Whereas, $m_H = (m_1 + m_2)/2$ is the average mass so that $m_C \leq m_H$ which is true in general. From band theory in optimally doped cuprates, $m_H \cong m_C \approx 1.7 m_0$ where $m_0$ is the free electron mass.[21] For optimally doped Bi2212, for example, we observe $m_H \cong 2.6 m_0$ with very little frequency dependence between 50 cm$^{-1}$ and 1000 cm$^{-1}$.[22] The IR Hall data on underdoped LSCO and YBCO imply a factor of three reduction in $m_H$ between optimal doping and p=0.10.[6,7] Moreover, if the pockets are elongated, as predicted by DW models for hole doped cuprates and observed in ARPES, a further reduction of $m_C$ relative to $m_H$ is expected.

Therefore there is a significant discrepancy between the observed $m_C$ in quantum oscillations and $m_H$ estimated for YBCO. This discrepancy might be accounted for by mass enhancement arising from interactions. The enhancement effects fall off with frequency so that it should be larger for DC than for the ~100 meV measurement frequencies of the IR Hall effect. Mass enhancement effects are also seen in $\sigma_{xx}$ and amount to a reduction in the effective mass by less than a factor of 2 at 100 meV compared with low frequencies for underdoped cuprates.[23] Moreover, $m_H$ in the IR Hall effect does not appear to be as much enhanced as m* in $\sigma_{xx}$.[22] Therefore, mass enhancement appears to be too small to account for the observed discrepancy between $m_H$ and $m_C$.

In general these quantum oscillation experiments makes Fermiology a high priority in the cuprates. Additional quantum oscillation measurements on other materials, particularly materials without chain bands would be highly desirable. There will be renewed interest in transport and optical measurements. Far IR Hall measurements in both electron and hole doped cuprates with and without chains is particularly interesting. IR Hall measurements at far IR frequencies ($\omega \leq 10\ meV$) would be below any gap energy and therefore would relate to properties at the Fermi surface. Also, at such low frequencies any mass enhancement effects should be present in $m_H$ making the comparison with $m_C$ more direct. The doping dependence of the Hall frequency can be compared with the cyclotron frequency from quantum oscillations, and with SDW models[8,13,14] and other theories of Fermi pockets. In addition, since IR Hall data allows the separation of band structure effects and quasiparticle scattering effects,[6,7,22,24] the temperature dependence will contribute to a better understanding of the anomalous temperature dependence of resistivity and Hall coefficient generally observed in magneto-transport measurements on the cuprates.[25]

Useful discussions with Sudip Chakravarty, Richard Greene, and Victor Yakovenko are gratefully acknowledged. Work was supported by NSF grant DMR-0303112.